\documentstyle[12pt]{article}
\newcommand{\be}{\begin{equation}}
\newcommand{\bea}{\begin{eqnarray}}
\newcommand{\eea}{\end{eqnarray}}
\newcommand{\ba}{\begin{array}}
\newcommand{\ea}{\end{array}}
\newcommand{\ee}{\end{equation}}

\expandafter\ifx\csname mathbbm\endcsname\relax

\else

\fi
\def\l{\label}
\def\o{\over}

\begin{document}
\begin{titlepage}
\hfill
\vbox{
    \halign{#\hfil         \cr
           hep-th/9802151 \cr
           IPM-98-269   \cr
           } 
      }  
\vspace*{3mm}
\begin{center}
{\LARGE On the Brane Configuration of $N=(4,4)$ 2D Supersymmetric 
Gauge Theories \\}
\vspace*{20mm}
{\ Mohsen Alishahiha \footnote{e-mail:alishah@physics.ipm.ac.ir}}\\
\vspace*{1mm}
{\it Institute for Studies in Theoretical Physics and Mathematics, \\
 P.O.Box 19395-1795, Tehran, Iran } \\
\vspace*{25mm}
\end{center}
\begin{abstract}
We study two dimensional $N=(4,4)$ supersymmetric gauge theories with 
various gauge groups and various hypermultiplets in the fundamental as 
well as bi-fundamental and adjoint representations. They have "mirror 
theories"  which become equivalent to them at strong coupling.
The theory with one fundamental and one adjoint
has a Higgs branch which is parametrized by the adjoint 
matter. We also consider theories which involve an orientifold plane. 
The brane realization of the Matrix theory formulation of NS 5-branes in 
Type II string theories is also considered.
\end{abstract}
\end{titlepage}
\newpage

\section{ Introduction}
Recently, supersymmetric gauge theory in various dimensions with various
supercharges have been studied in the context of brane theory. There are two
approaches for this purpose. One is considering wrapped D-branes on the 
Calabi-Yau cycles in Type II A and B string theories compactified on 
K3-fibred Calabi-Yau 3-fold \cite{KV} (For review see \cite{K}).
This approach was generalized in \cite{GEO} to explain mirror symmetry
in $N=4$ three dimensional gauge theory as well as Seiberg-Witten models
both with simple groups and product of simple gauge groups.

Another approach is to consider the configuration of intersecting 
D-branes and
NS 5-branes in Type II A and B string theories. At first, Hanany and Witten
\cite{HW} used a particular brane configuration in Type II B string theory 
in order to describe mirror symmetry in three dimensional supersymmetric
gauge theories with 8 supercharges \cite{IS}. Applying string duality to the 
configuration of branes in Type II B string theory, one can provide an
explanation for the mirror symmetry.

It has been shown \cite{EGK} by a particular brane configuration of type 
II A string
theory that one can obtain the supersymmetric $U(n)$ gauge theory in four 
dimension
with four supercharges. Then by making a certain deformation in the brane 
configuration, Seiberg's duality can be realized.
Introducing an orientifold plane in the brane configuration helps us to 
generalize this work to other classical Lie gauge groups. \cite{SO}

$N=2$ four dimensional gauge theory with gauge group $SU(n)$ was also 
obtained from the brane configuration in type II A string theory\cite{WIT}.
By lifting from type II A to M-theory, the exact solution
of $N=2$ $D=4$ SYM theory was obtained. More precisely, from the M-theory 
point of view, we have a five brane with worldvolume $R^{3,1} \times 
\Sigma$; where the 
theory on the $R^{3,1}$ is an $N=2$ $D=4$ SYM theory; and $\Sigma$ is the
Seiberg-Witten curve corresponding to it. This work was generalized to other
classical Lie gauge groups in \cite{GSU}.

In this context, chiral gauge theories in four dimensions with 4 supercharges
have been obtained by studying the brane configurations in the non-flat 
spacetime backgrounds, specially in the orbifold background \cite{T}. 
These theories have been also studied by introducing an orientifold six 
plane.

Recently, two dimensional $N=(4,4)$ supersymmetric gauge theories have been
studied by considering a particular brane configuration in Type II A 
string theory \cite{BA}. The same configuration was used in \cite{IN} to
show the relation between the Higgs branch of the theory and the moduli space
of the instantons \cite{WIT2}. 
similar configuration is considered in\cite{AHA}. Two dimensional 
$N=(2,2)$ supersymmetric gauge theories as well as five and six dimensional
theories have been also studied in \cite{HH},\cite{G6}\cite{G5}.
( For review of this approach see \cite{RE}.)

In this article we consider a class of brane configurations of Type II A
string theory to study several two dimensional gauge theories which have
8 or 16 supercharges. These theories have been appeared in the context of 
Matrix theory description of NS 5-brane in Tpye II A and B \cite{SS}. 
It has been shown that coincident NS 5-branes in Type II A and B lead to
non-trivial gauge theory in the limit of $g_{A,B} \rightarrow 0$ \cite{S}.
In \cite{WIT3} Witten obtained new gauge theories which can be described
by $ (p,q) $ 5-branes in Type II B or equivalently by considering Type II A/
M-theory on the non-flat background. He was also shown that the Matrix theory
formulation of these theories are two dimensional supersymmetric gauge theory
with 8 supercharges.

In section two we introduce our brane configuration in Type II A
string theory. We will consider the theory with one fundamental and one
adjoint matter. This model has a Higgs branch, but there is no smooth
transition from the Coulomb branch to the Higgs branch in the week
coupling limit. This transition can be seen by going to the strong
coupling limit. We also consider the brane configuration in the 
presence of an O 4-plane. So we can study the theory with $SP(N)$
gauge group and $D_{N_f}$ singularity. In section three we lift 
this brane configuration to
M-theory. We will see that there are two theories which become
equivalent in the strong coupling limit. In section four we 
consider the brane realization of the Matrix theory formulation of 
NS 5-branes in Type II string theory. It seems that the mirror 
symmetry in two dimensions which we will study in section three 
corresponds to the duality between Type II A on an $A_{k-1}$
singularity and $k$ NS 5-branes in Type II B string theory.
Note that in \cite{GOMIS} the correspondence between M-theory 5-branes 
and ALE backgrounds in Type II B string theory is realized as three
dimensional mirror symmetry. This correspondence is studied by
the Matrix theory description of them. 

\section{ Brane Configuration}

The Type II A brane configuration which we will use, involve three kinds of
branes.

1) An NS 5-brane with worldvolume $(x^0,x^1,x^2,x^3,x^4,x^5)$ which lives at
a point in the $(x^6,x^7,x^8,x^9)$ ( and $x^{10}$ in M-theory point of 
view) directions. Let $Q_L$ and $Q_R$ be the left and right 
moving supercharges in Type II A. The supercharges obey: 
\be 
\Gamma^0...\Gamma^9 Q_L=Q_L, \,\,\,\,\, \Gamma^0...\Gamma^9 Q_R=-Q_R
\ee  
NS 5-brane is invariant under half of the supersymmetries $\epsilon_L Q_l+
\epsilon_R Q_R$ with:
\be  \l{NS}
\Gamma^0...\Gamma^5 \epsilon_L=\epsilon_L, \,\,\,\,\, 
\Gamma^0...\Gamma^5 \epsilon_R=\epsilon_R
\ee

2) A D 2-brane with worldvolume $(x^0,x^1,x^6)$ at the point 
$(x^2,x^3,x^4,x^5,
x^7,x^8,x^9)$. It is invariant under half of the supersymmetries
\be   \l{D2}
\Gamma^0\Gamma^1\Gamma^6 \epsilon_R=\epsilon_L
\ee

3) A D 4-brane with worldvolume $(x^0,x^1,x^7,x^8,x^9)$ living at the
point $(x^2,x^3,x^4$, $x^5, x^6)$ and invariant under half of the 
supersymmetries with
\be   \l{D4}
\Gamma^0\Gamma^1\Gamma^7\Gamma^8\Gamma^9 \epsilon_R=\epsilon_L
\ee

Each relation (2), (3), (4) by itself breaks half of
the sypersymmetries, but it is easy to see that they are not independent, so
altogether they break ${1 \o 4}$ of the original supersymmetries of Type II A
string theory.

The presence of all these branes break the ten dimensional Lorentz group
to 1+1 dimensional Lorentz group with $SO(4)\times SU(2)_R$ global symmetry.
$SO(4)$ corresponds to rotation in the $ x^2,x^3,x^4,x^5 $ directions and 
$SU(2)_{R}$ to the rotation in the $x^7,x^8,x^9$ directions.
Brane configuration which we will consider, consists of NS 5-branes at
the points $r_i=(x^7_i,x^8_i,x^9_i)$ and $x^6_i$; the D 2-branes 
which are suspended 
between these NS 5-branes; so that they are finite in $x^6$ direction. 
Also we
will have some D 4-branes at $m_i=(x^2_i,x^3_i,x^4_i,x^5_i)$ in between 
the NS 5-branes. This configuration of branes preserves 8 supercharges in 
intersection worldvolume $(x^0,x^1)$. So we have an $N=(4,4)$ gauge theory in
two dimensions. 

For example consider two NS 5-branes at $r_1,x^6_1=0$ and
$r_2,x^6_2=L$, and N D 2-branes suspended between them, and $N_f$ D 4-branes
in between these two NS 5-branes at points $m_i, x^6_i$. 

\vspace*{1mm}
\begin{center}
\unitlength 1mm
\linethickness{0.4pt}
\begin{picture}(85.00,49.00)
\put(23.00,20.00){\line(0,1){25.00}}
\put(42.00,45.00){\line(0,-1){25.00}}
\put(31.67,25.00){\line(-1,-1){9.33}}
\put(34.67,24.67){\line(-1,-1){8.67}}
\put(38.00,24.67){\line(-1,-1){8.33}}
\put(29.67,16.33){\line(0,0){0.00}}
\put(23.00,40.33){\line(1,0){19.00}}
\put(42.00,37.67){\line(-1,0){19.00}}
\put(23.00,34.67){\line(1,0){19.00}}
\put(65.00,30.33){\vector(0,1){14.00}}
\put(65.00,30.67){\vector(-1,-1){9.67}}
\put(65.00,30.33){\vector(1,0){14.67}}
\put(64.00,49.00){$x^2,x^3,x^4,x^5$}
\put(85.00,30.33){$x^6$}
\put(54.33,16.67){$x^7,x^8,x^9$}
\put(35.67,8.67){Notation}
\end{picture}
\end{center}
\vspace*{.4mm}

$r=r_2-r_1$ is the Fayet-Iliopoules terms and is in $(1,1,3)$ representation 
of $SO(4)\times SU(2)_{R}$, where the representation of $SO(4)$ are 
labelled by $SU(2)\times SU(2)$. When $r=0$, the D 2-branes can suspend 
between two NS 5-branes and 
preserve 8 supercharges in 1+1 dimension. In this case we have a $U(N)$ gauge
theory which is in the Coulomb phase and can be parametrized by scalars 
in vector
multiplet. In brane language, these scalars are fluctuations of D 2-branes in
$x^2,x^3,x^4,x^5$ directions; we set $u=x^2+ix^3, v=x^4+ix^5$. They 
transform as $(2,2,1)$ of $SO(4)\times SU(2)_R$.

The presence of  $N_f$ D 4-branes between the two NS 5-branes 
correspond to $N_f$ matter in the fundamental representation of the gauge 
groups. In fact, strings stretched between D 2-branes and D 4-branes give us
hypermultiplets in the fundamental representation. The position of these
D 4-branes, $m_i$, are bare masses of the hypermultiplets which transform as  
$(2,2,1)$ under global symmetry $SO(4)\times SU(2)_R$. If two of $m_i$'s 
become equal, the D 2-brane can break and suspends between these two D 
4-branes. 
In this case the scalars in the hypermultiplets correspond to the fluctuations
of D 2-branes in the directions $x^7,x^8,x^9$ and also one should include the 
component $A_6$ from the gauge field which can survive from boundary 
conditions. These scalars parametrize the Higgs branch of the theory.
So, along the Higgs branch there are scalars transforming in $3+1$ of 
$SU(2)_R$ \cite{DS}.
If we look at this configuration from the point 
of view of M-theory the last one which is singlet under $SU(2)_R$, 
becomes more manifest; in fact it is the fluctuations in $x^{10}$ (compact 
direction in M-theory). The position of NS 5-branes in $x^{10}$ 
can be interpreted as $\theta$ angle ($\theta=x^{10}_2-x^{10}_1$)

The distance between two NS 5-branes determines the gauge coupling constant 
of the two dimensional theory; more precisely
\be
{1 \o g_2^2}={L \o g }
\ee
where $g$ is the string coupling. The quantum Coulomb branch can occur when 
$r=0$ and $\theta=0$ \cite{SOWI}. 
If $r\neq0$ (or $\theta \neq 0$) the theory can not have the Coulomb branch, 
which means that if we want to have a supersymmetric configuration, 
the D 2-branes must be broken into D 2-branes between NS 5-branes and 
D 4-branes. Note that in this case one should also consider that 
the s-configuration is not supersymmetric \cite{HH}. 
If the D 4-branes have the same position in $x^2,x^3,x^4,x^5$ 
(equal mass $m_i=m_j$), the  theory can be in the Higgs branch. 
The complete Higgsing is only possible for $2N \leq N_f$.

If we have several NS 5-branes one can also have another hypermultiplet.
Strings between D 2-branes which end from left and right to a NS 5-brane 
give us bi-fundamental matter. In particular if we compact the $x^6$
direction we can also have hypermultiplets in the adjoint representation.
This can give us a two dimensional gauge theory with 16 supercharges 
\cite{IN}.
 
An interesting model is $U(1)$ gauge theory with $N_f=1$ and one hypermultipet 
in the adjoint representation as $U(1)$ is Abelian is simply free. This 
theory has been recently studied and argued that it can have a Higgs branch 
\cite{WIT2}. In brane language
it is easy to see the occurrence of the Higgs branch. In this case we have
one NS 5-brane and one D 2-brane which wraps around $x^6$ direction and 
both ends are in the NS 5-brane ( in fact it intersects the NS 5-brane). 
We also have a D 4-brane. The D 2-brane can end 
on the NS 5-brane and move in $x^2,x^3,x^4,x^5$, so the theory is in the 
Coulomb branch. When D 2-brane intersects D 4-brane it can break into two
D 2-branes between D 4-brane and NS 5-brane. One can also imagine that
the position of these two D 2-branes in the directions of $x^2,x^3,x^4,x^5$
become equal, so they can connect and leave the NS 5-brane. This means that
the adjoint hypermultiplet (free hypermultiplet) gets expectation value and 
the theory is in the
Higgs branch, which is parametrized by the position of the D 2-brane in the
directions $x^7,x^8,x^9$ and $x^6$ ($A_6$ component of gauge field). So the
model has a Higgs phase as well as a Coulomb phase, but in this case the
Higgs branch is obtained from the expectation value of matter in the adjoint 
representation as studied in \cite{WIT2}. 
\vspace*{2mm}  
\begin{figure} 
\unitlength 1mm
\linethickness{0.4pt}
\begin{picture}(115.00,88.00)
\multiput(40.00,75.00)(0.99,-0.10){3}{\line(1,0){0.99}}
\multiput(42.97,74.70)(0.36,-0.11){8}{\line(1,0){0.36}}
\multiput(45.83,73.82)(0.22,-0.12){12}{\line(1,0){0.22}}
\multiput(48.45,72.39)(0.13,-0.11){17}{\line(1,0){0.13}}
\multiput(50.74,70.47)(0.12,-0.15){16}{\line(0,-1){0.15}}
\multiput(52.60,68.14)(0.11,-0.22){12}{\line(0,-1){0.22}}
\multiput(53.96,65.48)(0.12,-0.41){7}{\line(0,-1){0.41}}
\multiput(54.77,62.60)(0.11,-1.49){2}{\line(0,-1){1.49}}
\multiput(55.00,59.63)(-0.09,-0.74){4}{\line(0,-1){0.74}}
\multiput(54.62,56.66)(-0.12,-0.35){8}{\line(0,-1){0.35}}
\multiput(53.67,53.83)(-0.11,-0.20){13}{\line(0,-1){0.20}}
\multiput(52.18,51.24)(-0.12,-0.13){17}{\line(0,-1){0.13}}
\multiput(50.20,49.00)(-0.15,-0.11){16}{\line(-1,0){0.15}}
\multiput(47.82,47.20)(-0.24,-0.12){11}{\line(-1,0){0.24}}
\multiput(45.13,45.90)(-0.41,-0.11){7}{\line(-1,0){0.41}}
\multiput(42.24,45.17)(-1.49,-0.07){2}{\line(-1,0){1.49}}
\multiput(39.25,45.02)(-0.74,0.11){4}{\line(-1,0){0.74}}
\multiput(36.30,45.46)(-0.31,0.11){9}{\line(-1,0){0.31}}
\multiput(33.49,46.49)(-0.20,0.12){13}{\line(-1,0){0.20}}
\multiput(30.94,48.04)(-0.13,0.12){17}{\line(-1,0){0.13}}
\multiput(28.75,50.07)(-0.12,0.16){15}{\line(0,1){0.16}}
\multiput(27.01,52.50)(-0.11,0.25){11}{\line(0,1){0.25}}
\multiput(25.78,55.22)(-0.11,0.49){6}{\line(0,1){0.49}}
\put(25.12,58.13){\line(0,1){2.99}}
\multiput(25.04,61.12)(0.10,0.59){5}{\line(0,1){0.59}}
\multiput(25.56,64.06)(0.11,0.28){10}{\line(0,1){0.28}}
\multiput(26.65,66.84)(0.12,0.18){14}{\line(0,1){0.18}}
\multiput(28.27,69.35)(0.12,0.12){18}{\line(0,1){0.12}}
\multiput(30.36,71.49)(0.16,0.11){15}{\line(1,0){0.16}}
\multiput(32.83,73.17)(0.28,0.12){10}{\line(1,0){0.28}}
\multiput(35.58,74.33)(0.74,0.11){6}{\line(1,0){0.74}}
\multiput(100.00,75.00)(0.99,-0.10){3}{\line(1,0){0.99}}
\multiput(102.97,74.70)(0.36,-0.11){8}{\line(1,0){0.36}}
\multiput(105.83,73.82)(0.22,-0.12){12}{\line(1,0){0.22}}
\multiput(108.45,72.39)(0.13,-0.11){17}{\line(1,0){0.13}}
\multiput(110.74,70.47)(0.12,-0.15){16}{\line(0,-1){0.15}}
\multiput(112.60,68.14)(0.11,-0.22){12}{\line(0,-1){0.22}}
\multiput(113.96,65.48)(0.12,-0.41){7}{\line(0,-1){0.41}}
\multiput(114.77,62.60)(0.11,-1.49){2}{\line(0,-1){1.49}}
\multiput(115.00,59.63)(-0.09,-0.74){4}{\line(0,-1){0.74}}
\multiput(114.62,56.66)(-0.12,-0.35){8}{\line(0,-1){0.35}}
\multiput(113.67,53.83)(-0.11,-0.20){13}{\line(0,-1){0.20}}
\multiput(112.18,51.24)(-0.12,-0.13){17}{\line(0,-1){0.13}}
\multiput(110.20,49.00)(-0.15,-0.11){16}{\line(-1,0){0.15}}
\multiput(107.82,47.20)(-0.24,-0.12){11}{\line(-1,0){0.24}}
\multiput(105.13,45.90)(-0.41,-0.11){7}{\line(-1,0){0.41}}
\multiput(102.24,45.17)(-1.49,-0.07){2}{\line(-1,0){1.49}}
\multiput(99.25,45.02)(-0.74,0.11){4}{\line(-1,0){0.74}}
\multiput(96.30,45.46)(-0.31,0.11){9}{\line(-1,0){0.31}}
\multiput(93.49,46.49)(-0.20,0.12){13}{\line(-1,0){0.20}}
\multiput(90.94,48.04)(-0.13,0.12){17}{\line(-1,0){0.13}}
\multiput(88.75,50.07)(-0.12,0.16){15}{\line(0,1){0.16}}
\multiput(87.01,52.50)(-0.11,0.25){11}{\line(0,1){0.25}}
\multiput(85.78,55.22)(-0.11,0.49){6}{\line(0,1){0.49}}
\put(85.12,58.13){\line(0,1){2.99}}
\multiput(85.04,61.12)(0.10,0.59){5}{\line(0,1){0.59}}
\multiput(85.56,64.06)(0.11,0.28){10}{\line(0,1){0.28}}
\multiput(86.65,66.84)(0.12,0.18){14}{\line(0,1){0.18}}
\multiput(88.27,69.35)(0.12,0.12){18}{\line(0,1){0.12}}
\multiput(90.36,71.49)(0.16,0.11){15}{\line(1,0){0.16}}
\multiput(92.83,73.17)(0.28,0.12){10}{\line(1,0){0.28}}
\multiput(95.58,74.33)(0.74,0.11){6}{\line(1,0){0.74}}
\put(40.00,65.00){\line(0,1){20.00}}
\put(100.00,85.00){\line(0,-1){2.33}}
\put(100.00,81.67){\line(0,-1){2.33}}
\put(100.00,78.33){\line(0,-1){2.33}}
\put(100.00,75.00){\line(0,-1){2.33}}
\put(100.00,71.67){\line(0,-1){2.33}}
\put(100.00,68.33){\line(0,-1){2.33}}
\put(105.33,52.00){\line(-1,-1){13.33}}
\put(46.00,51.67){\line(-1,-1){3.67}}
\put(41.00,46.67){\line(-1,-1){3.67}}
\put(36.33,42.00){\line(-1,-1){3.67}}
\put(35.33,88.00){NS 5-brane}
\put(96.67,87.67){NS 5-brane}
\put(89.00,35.67){D 4-brane}
\put(29.33,35.67){D 4-brane}
\put(29.33,26.00){Coulomb branch}
\put(89.00,26.00){Higgs branch}
\end{picture}
\end{figure} 

For gauge group $U(N)$ the story is the same as $U(1)$. The Higgs branch 
is parametrized by motion of ends of D 2-branes in the directions 
$x^7,x^8,x^9$. This can happen if D 2-branes leave NS 5-brane.
In this case the Higgs branch is also parametrized by the adjoint 
hypermultiplets. In the Higgs branch we have N D 2-branes that end on the
D 4-brane which can be interpreted by N points in the $R^4$. Of course 
one should mod the 
Weyl group action on them. So the Higgs branch of the theory is the symmetric
product of the Higgs branch of N=1, i.e. $S^N R^4$ \cite{WIT2}.

The Coulomb branch can occur when D 2-branes end on the NS 5-brane and move
in the directions $x^2,x^3,x^4,x^5$. In this case the gauge group is broken
to $U(1)^N$. Note that although it is difficult to see the transition from
Coulomb branch to Higgs branch in the context of brane configuration, it
is easy to see that the final configuration can occur; which means 
that there is Higgs branch for the case $N_f=1$. In the next section we 
will see that this
transition can happen by going to strong coupling limit.
Note also that as we will see, in the strong 
coupling limit of the theory, these two branch become equivalent.

One can also add an orientifold plane parallel to D 4-branes in the Type
II A brane configuration. Consider an O 4-plane parallel to D 4-branes with
worldvolume $(x^0,x^1,x^7,x^8,x^9)$. The orientifold plane can be introduced
by moding out $(x^2,x^3,x^4,x^5,$ $x^6)\rightarrow 
(-x^2,-x^3,-x^4,-x^5,-x^6)$
together with gauging of the world sheet parity. There are two type of 
O 4-plane in the Type II A 
String theory, classified with the RR charges \cite{RR}. One type has -1 
D 4-brane charge and the other has +1. When 2N D 4-branes are close to the 
fixed
point, the former leads to $SO(2N)$ gauge enhancement and the latter to
$SP(2N)$.
Since we are interested in two dimensional theory, this enhancement 
corresponds to global symmetry. 

Now consider the following brane configuration. 2 NS 5-branes, 2 D 2-branes
(in fact only one half of them is physical the other is its image 
with respect to
O 4-plane), $2N_f$ D 4-branes and an O 4-plane parallel to them. We choose
O 4-plane charge to be negative of that of the D 4-brane, so the 
global symmetry is 
$SO(2N_f)\times SO(4) \times SU(2)_R$. In the presence of the O 4-plane 
the Coulomb branch is $R^4/Z_2$ and the gauge group is $SP(1)\simeq SU(2)$.
This theory is the $D_{N_f}$ model discussed in \cite{DS}.

In the spirit of \cite{HW}, presence of $N_f$ D 4-branes induce magnetic charge
in the $(u,v)$ space. Note that the O 4-plane also induces twice magnetic 
charge with opposite sign. They can affect the metric on the Coulomb branch.
Minimizing the total NS 5-brane worldvolume, we find the four dimensional 
Laplace equation; therefore $SU(2)$ gauge symmetry with $2N_f$ flavours 
has the following metric on the Coulomb branch.
\be
ds^2=({1\o g^2}+{{2N_f-2} \o X^2})d^2X
\ee
where $X^2=u{\bar u}+v{\bar v}$. In the case of $m_i \neq0$ we find
\be
ds^2=({1\o g^2}+\sum_{i=1}^{N_f}{1 \o |X-m_i|^2} +{1 \o |X+m_i|^2}-
{2 \o X^2}) d^2X 
\ee
which is the one-loop correction to the metric in the Coulomb 
branch\cite{DS}.
Note that the last term is the effect of O 4-plane. As in \cite{BA}, the 
torsion on the Coulomb branch can be interpreted as the anti-symmetric 
$B_{\mu,\nu}$ living on the NS 5-brane of Type II A string theory. In our case  
the B-field charge is $2N_f-2$. Note that in the case $N_f=1$ the effect
of O 4-plane can cancel by these two D 4-branes, so the metric will be well
defined over all the range of the moduli space. For $N_f=0$ there is only O 
4-plane, so there is no global symmetry as well as the Higgs branch. Also 
metric is not flat and not Riemannian and it has a singularity at finite 
distance $X=\sqrt{2}g$.

From the above discussion, it seems that for N D 2-branes and $N_f$ D 
4-branes (and one should also consider their images) the theory is $SP(N)$ 
with $D_{N_f}$ singularity and the Coulomb branch is $(R^4)^N/W$, where 
$W$ is Weyl group of $SP(N)$.

\section{M-theory description and mirror symmetry}

The brane configuration in M-theory consists of M 5-branes and M 
2-branes. By lifting to M-theory the D 4-branes become M 5-branes. In this case the
Lorentz group of 11-dimensional M-theory is broken to
$SO(1,1)\times SO(4)\times SO(4)$ by our brane configuration. These two
$SO(4)$'s can be interpreted as R-symmetry. The first $SO(4)$ which acts 
on the
$x^2,x^3,x^4,x^5$ directions, corresponds to R-symmetry of the Higgs branch.
The second $SO(4)$ which acts on the $x^7,x^8,x^9,x^{10}$ directions, is the
R-symmetry of the Coulomb branch. Note that the R-symmetry of the Coulomb
branch of the theory is enhanced from $SU(2)_R$ to $SO(4)$ at strong coupling
\cite{BA}, as
conjectured in \cite{WIT2}. So in the strong coupling, the $R$-symmetry of
the Higgs and Coulomb branches are the same. It is very similar to the three
dimensional $N=4$ gauge theory, where we have mirror symmetry,
which exchanges Coulomb
and Higgs branches. Therefore we expect to see a similar 
symmetry in two dimensional $N=(4,4)$ gauge theory in strong coupling. 
In fact there are two different theories that become equivalent at strong
coupling. More precisely, the Coulomb branch of one theory is equivalent to
the Higgs branch of the other theory at strong coupling. When we are in 
strong coupling limit - M-theory - there is no difference between these 
theories. In brane language it means that, they are the same brane 
configurations in M-theory.
If we have a brane configuration in M-theory, the theory which we will find
in Type II A limit, depends on the direction compactified.

Consider an operator $U$ acting as $x^i \rightarrow x^{i+5}$ and
$x^{i+5} \rightarrow -x^i$ for $i=2,3,4,5$. $U$ is a reflection 
operator which changes the theory to its mirror. Assume we have a 
particular brane
configuration in M-theory. It can be obtained from a brane configuration of 
Type II A by opening up the $11^{th}$ direction. Now by applying the operator
$U$ on this configuration; then going back to the Type II A, we will find 
the mirror theory which is equivalent to the first one we 
began with in the strong coupling limit. Note that under $U$, 
the number of degrees of freedom of the theories do not change, so the 
superconformal field theories which we will obtain from these two, one from 
the Higgs branch and another from the Coulomb branch, will have the same 
central charge.

If we obtain our brane configuration from T-dualizing 
the brane configuration of Type II B, which describes the mirror symmetry 
of $N=4$, $D=3$ sypersymmetric gauge theory \cite{HW}, then the 
$U$-transformation will be  equivalent to $RS$ transformation introduced in 
\cite{HW}. In fact exchanging of $x^2,x^3,x^4$ with $x^7,x^8,x^9$ 
corresponds to the $R$ transformation and exchanging $x^5$ with $x^{10}$ 
corresponds to $S$-duality. 

As an example, consider the theory with gauge group $U(N)$ and $N_f$  
hypermultiplets in the fundamental representation \footnote{ Brane 
realization of this theory helps us to write the metric of the
moduli space of the Coulomb branch for gauge groups with arbitrary
rank\cite{SA}}. It has the following 
brane configuration. (for simplicity the case N=3, $N_f=7$ is indicated)
\vspace*{2mm}   
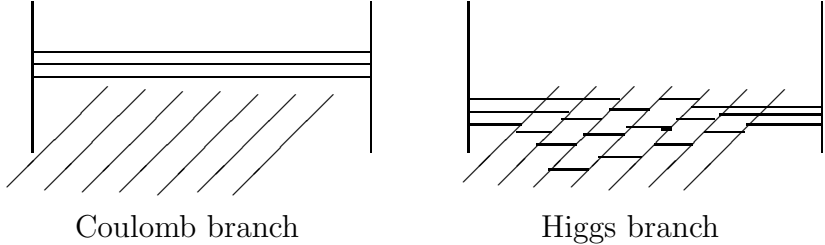
\begin{center}
\unitlength 1mm
\linethickness{0.4pt}
\begin{picture}(120.00,40.00)
\put(15.00,20.00){\line(0,1){20.00}}
\put(25.00,28.67){\line(-1,-1){13.33}}
\put(30.00,28.33){\line(-1,-1){13.33}}
\put(35.00,28.00){\line(-1,-1){13.33}}
\put(40.00,28.00){\line(-1,-1){13.33}}
\put(45.00,27.67){\line(-1,-1){13.33}}
\put(50.00,27.67){\line(-1,-1){13.00}}
\put(55.00,27.67){\line(-1,-1){13.33}}
\put(60.00,20.00){\line(0,1){20.00}}
\put(60.00,30.00){\line(-1,0){45.00}}
\put(15.00,31.67){\line(1,0){45.00}}
\put(15.00,33.33){\line(1,0){45.00}}
\put(85.00,28.67){\line(-1,-1){12.67}}
\put(90.00,28.67){\line(-1,-1){13.00}}
\put(95.00,28.67){\line(-1,-1){13.33}}
\put(100.00,28.67){\line(-1,-1){13.33}}
\put(105.00,28.33){\line(-1,-1){13.33}}
\put(110.00,28.33){\line(-1,-1){13.00}}
\put(115.00,28.33){\line(-1,-1){13.33}}
\put(79.33,22.67){\line(1,0){4.67}}
\put(104.33,22.67){\line(1,0){5.33}}
\put(105.67,24.33){\line(-1,0){5.00}}
\put(90.67,24.33){\line(-1,0){5.33}}
\put(82.00,21.00){\line(1,0){5.33}}
\put(97.67,21.00){\line(1,0){5.00}}
\put(99.00,23.33){\line(-1,0){4.67}}
\put(98.67,23.00){\line(1,0){1.33}}
\put(100.00,23.33){\line(-1,0){6.00}}
\put(99.00,23.33){\line(-1,0){5.00}}
\put(93.67,22.33){\line(-1,0){5.33}}
\put(91.67,25.67){\line(1,0){5.33}}
\put(98.33,27.00){\line(1,0){5.33}}
\put(96.00,19.33){\line(-1,0){5.67}}
\put(89.00,17.67){\line(-1,0){5.33}}
\put(73.00,20.00){\line(0,1){20.00}}
\put(120.00,40.00){\line(0,-1){20.00}}
\put(120.00,26.00){\line(-1,0){17.33}}
\put(106.33,25.00){\line(1,0){13.67}}
\put(120.00,23.67){\line(-1,0){10.00}}
\put(93.00,27.00){\line(-1,0){20.00}}
\put(73.00,25.33){\line(1,0){13.33}}
\put(80.00,23.67){\line(-1,0){7.00}}
\put(20.67,8.67){Coulomb branch }
\put(82.67,8.67){Higgs branch }
\end{picture}
\end{center}
\vspace*{.1mm}   

Since the Higgs branch of the theory is equivalent to the Coulomb branch of 
the other theory in the strong coupling, the first theory should be in the 
complete Higgs branch, so $2N\leq N_f$. To finding the mirror theory,
one should first Higgs the theory, then go to the M-theory and after 
$U$-transforming go backing to the Type II A. Doing so, we will find 
the theory with gauge group: 
$U(1) \times U(2) \times \cdots U(N-1) \times U(N)^{N_f-2N+1} \times U(N-1)
\cdots \times U(1)$ with hypermultiplets transforming as
$(1, 2) \oplus (2,{\bar 3}) \oplus \cdots (N-1,{\bar N}) \oplus
N \oplus (N,{\bar N}) \oplus \cdots (N,{\bar N}) \oplus N \oplus
(N, {\bar {N-1}}) \oplus \cdots (2,1)$. In finding  the matter content of 
the mirror theory we used the fact that the $x^6$ coordinate of D 4-branes
(matter) appear to be irrelevant as was noted in \cite{HW}. Although $x^6$ 
direction of matter has no physical meaning
in the first theory, it becomes physical in the mirror theory. In fact,
it becomes the gauge coupling in the mirror theory. As this is a 
strong coupling phenomena, it would be difficult to explain it in the 
context of field theory.

\vspace*{2mm}  
\begin{center}
\unitlength 1mm
\linethickness{0.4pt}
\begin{picture}(80.00,40.00)
\put(20.00,20.00){\line(0,1){20.00}}
\put(30.00,40.00){\line(0,-1){20.00}}
\put(40.00,20.00){\line(0,1){20.00}}
\put(50.00,40.00){\line(0,-1){20.00}}
\put(60.00,20.00){\line(0,1){20.00}}
\put(70.00,40.00){\line(0,-1){20.00}}
\put(80.00,20.00){\line(0,1){19.67}}
\put(20.00,30.67){\line(1,0){10.00}}
\put(30.00,33.00){\line(1,0){10.00}}
\put(40.00,28.33){\line(-1,0){10.00}}
\put(40.00,34.33){\line(1,0){10.00}}
\put(50.00,30.67){\line(-1,0){10.00}}
\put(40.00,26.67){\line(1,0){10.00}}
\put(50.00,35.67){\line(1,0){10.00}}
\put(60.00,32.33){\line(-1,0){10.00}}
\put(50.00,28.33){\line(1,0){10.00}}
\put(60.00,33.33){\line(1,0){10.00}}
\put(70.00,29.33){\line(-1,0){10.00}}
\put(70.00,31.00){\line(1,0){10.00}}
\put(58.00,25.00){\line(-1,-1){10.33}}
\put(47.67,25.00){\line(-1,-1){10.67}}
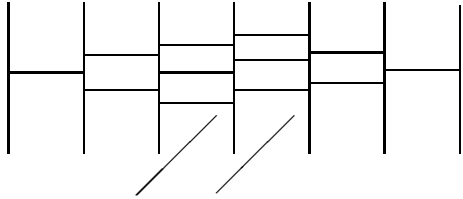
\put(17.33,6.33){The Coulomb branch of the mirror theory}
\end{picture}
\end{center}
\vspace*{.1mm}  

Note that under this symmetry, the number of bosonic degrees of freedom of 
the Higgs branch of the first theory (scalars in hypermultiplets) is equal 
to that of the number of
bosonic degrees of freedom of the Coulomb branch of the second one (scalars
in vector multiplets); so the supercoformal theories corresponding to them
have the same central charges. 
One can also start with the Coulomb branch of the first theory. In this case
after $U$-transformation we will find the Higgs branch of the second one.
  
Let us return to the case $N_f=1$ with one adjoint hypermultiplet. If we
begin with the Coulomb branch of the theory and apply the 
$U$-transformation, we will end up with the theory in the Higgs branch. So 
as we
said, the transition from Coulomb branch to Higgs branch can occur when 
we go to the strong coupling and apply $U$-transformation.
Note that in the M-theory limit, there is no difference between these two
brane configuration, which means the Coulomb branch and the Higgs branch
of the theory become equivalent at strong coupling.

By introducing the $U$-transformation it is possible to study a large class
of theories and their mirrors. In principal we can start by an arbitrary
brane configuration in the Type II A string theory (in Higgs branch or
Coulomb branch or a mixed branch) then lift to the M-theory an apply
$U$-transformation, then back to the Type II A. In this case we can find 
a different theory which becomes equivalent in strong coupling limit, 
to the original theory; Although it may not have a Lagrangian formalism.

Consider the model (A-model) which has $U(N)$ gauge group, $N_f$ 
hypermultiplets in the fundamental representation of the gauge group, and one
hypermultiplet in the adjoint representation. In the brane language, this model
consists of a NS 5-brane, N D 2-branes which wrap on the $x^6$ direction 
and 
end on the NS 5-brane (in fact it can be viewed as intersecting), so there
is a hypermultiplet in the adjoint representation. We should also add $N_f$
D 4-branes at points $m_i$ and $x^6$ as hypermultiplets in the 
fundamental representation. In this model $x^6$ component of the matter 
appears to be irrelevant, but $m_i$'s ($\eta_i=m_i-m_{i+1}$) correspond to 
the bare masses of the hypermultiplets. 
\vspace*{2cm}
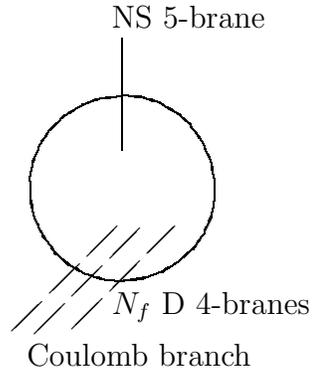
\begin{center}
\unitlength 1mm
\linethickness{0.4pt}
\begin{picture}(81.68,51.33)
\multiput(69.33,42.35)(0.85,-0.09){3}{\line(1,0){0.85}}
\multiput(71.89,42.08)(0.35,-0.11){7}{\line(1,0){0.35}}
\multiput(74.33,41.30)(0.20,-0.12){11}{\line(1,0){0.20}}
\multiput(76.56,40.02)(0.13,-0.11){15}{\line(1,0){0.13}}
\multiput(78.47,38.31)(0.12,-0.16){13}{\line(0,-1){0.16}}
\multiput(79.99,36.24)(0.12,-0.26){9}{\line(0,-1){0.26}}
\multiput(81.05,33.90)(0.11,-0.50){5}{\line(0,-1){0.50}}
\put(81.61,31.40){\line(0,-1){2.57}}
\multiput(81.63,28.83)(-0.10,-0.50){5}{\line(0,-1){0.50}}
\multiput(81.12,26.31)(-0.11,-0.26){9}{\line(0,-1){0.26}}
\multiput(80.10,23.96)(-0.11,-0.16){13}{\line(0,-1){0.16}}
\multiput(78.62,21.86)(-0.13,-0.12){15}{\line(-1,0){0.13}}
\multiput(76.74,20.11)(-0.20,-0.12){11}{\line(-1,0){0.20}}
\multiput(74.54,18.80)(-0.35,-0.12){7}{\line(-1,0){0.35}}
\multiput(72.11,17.96)(-0.85,-0.10){3}{\line(-1,0){0.85}}
\multiput(69.56,17.65)(-1.28,0.11){2}{\line(-1,0){1.28}}
\multiput(67.00,17.87)(-0.35,0.11){7}{\line(-1,0){0.35}}
\multiput(64.54,18.61)(-0.20,0.11){11}{\line(-1,0){0.20}}
\multiput(62.29,19.85)(-0.14,0.12){14}{\line(-1,0){0.14}}
\multiput(60.35,21.52)(-0.12,0.16){13}{\line(0,1){0.16}}
\multiput(58.79,23.57)(-0.11,0.23){10}{\line(0,1){0.23}}
\multiput(57.69,25.88)(-0.12,0.50){5}{\line(0,1){0.50}}
\put(57.09,28.38){\line(0,1){2.57}}
\multiput(57.02,30.95)(0.12,0.63){4}{\line(0,1){0.63}}
\multiput(57.48,33.47)(0.11,0.26){9}{\line(0,1){0.26}}
\multiput(58.45,35.85)(0.11,0.16){13}{\line(0,1){0.16}}
\multiput(59.90,37.97)(0.12,0.12){15}{\line(1,0){0.12}}
\multiput(61.75,39.75)(0.18,0.11){12}{\line(1,0){0.18}}
\multiput(63.93,41.11)(0.30,0.11){8}{\line(1,0){0.30}}
\multiput(66.34,41.98)(0.75,0.09){4}{\line(1,0){0.75}}
\put(69.33,35.00){\line(0,1){15.00}}
\put(76.33,25.00){\line(-1,-1){3.67}}
\put(71.67,20.33){\line(-1,-1){4.00}}
\put(66.67,15.33){\line(-1,-1){4.00}}
\put(71.67,24.67){\line(-1,-1){4.00}}
\put(66.67,19.67){\line(-1,-1){4.00}}
\put(61.67,14.67){\line(-1,-1){4.00}}
\put(68.67,25.00){\line(-1,-1){4.00}}
\put(63.67,20.00){\line(-1,-1){4.00}}
\put(58.67,15.00){\line(-1,-1){4.00}}
\put(68.00,51.33){NS 5-brane}
\put(68.00,13.33){$N_f$ D 4-branes}
\put(56.67,6.33){Coulomb branch}
\end{picture}
\end{center}

Assume the theory is in the Higgs branch, then the mirror model 
(B-model) will be the Coulomb branch of theory
with gauge group $\prod_{i=0}^{N_f-1} U(N)_i$ and its hypermultiplets consist
of one fundamental matter charged under $U(N)_1$ and $N_f$ bi-fundamental 
which have charge respectively under $U(N)_i\times U(N)_{i+1}$ in the
representation $(N, {\bar N})$ with the cyclic identification $i\sim i+N$.
\vspace*{2cm}
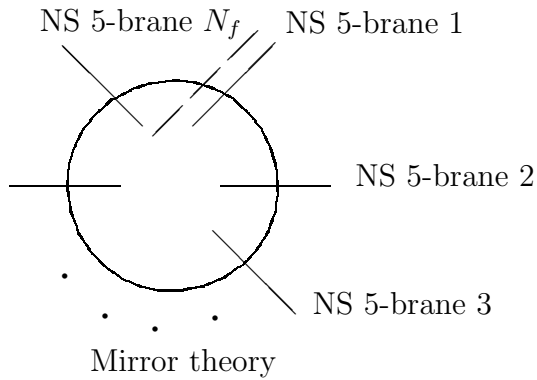
\begin{center}
\unitlength 1mm
\linethickness{0.4pt}
\begin{picture}(86.67,52.00)
\multiput(62.33,44.02)(0.94,-0.10){3}{\line(1,0){0.94}}
\multiput(65.15,43.73)(0.34,-0.11){8}{\line(1,0){0.34}}
\multiput(67.86,42.88)(0.21,-0.11){12}{\line(1,0){0.21}}
\multiput(70.34,41.50)(0.13,-0.12){16}{\line(1,0){0.13}}
\multiput(72.49,39.66)(0.12,-0.15){15}{\line(0,-1){0.15}}
\multiput(74.23,37.42)(0.11,-0.23){11}{\line(0,-1){0.23}}
\multiput(75.48,34.87)(0.12,-0.46){6}{\line(0,-1){0.46}}
\multiput(76.19,32.13)(0.07,-1.42){2}{\line(0,-1){1.42}}
\multiput(76.33,29.30)(-0.11,-0.70){4}{\line(0,-1){0.70}}
\multiput(75.90,26.49)(-0.11,-0.30){9}{\line(0,-1){0.30}}
\multiput(74.92,23.83)(-0.12,-0.19){13}{\line(0,-1){0.19}}
\multiput(73.42,21.43)(-0.11,-0.12){17}{\line(0,-1){0.12}}
\multiput(71.47,19.37)(-0.17,-0.12){14}{\line(-1,0){0.17}}
\multiput(69.14,17.75)(-0.26,-0.11){10}{\line(-1,0){0.26}}
\multiput(66.54,16.63)(-0.56,-0.11){5}{\line(-1,0){0.56}}
\put(63.76,16.06){\line(-1,0){2.84}}
\multiput(60.93,16.05)(-0.56,0.11){5}{\line(-1,0){0.56}}
\multiput(58.15,16.62)(-0.26,0.11){10}{\line(-1,0){0.26}}
\multiput(55.54,17.74)(-0.17,0.12){14}{\line(-1,0){0.17}}
\multiput(53.21,19.36)(-0.12,0.12){17}{\line(0,1){0.12}}
\multiput(51.26,21.41)(-0.12,0.18){13}{\line(0,1){0.18}}
\multiput(49.76,23.81)(-0.11,0.30){9}{\line(0,1){0.30}}
\multiput(48.77,26.47)(-0.11,0.70){4}{\line(0,1){0.70}}
\multiput(48.34,29.27)(0.07,1.42){2}{\line(0,1){1.42}}
\multiput(48.48,32.11)(0.12,0.46){6}{\line(0,1){0.46}}
\multiput(49.18,34.85)(0.11,0.23){11}{\line(0,1){0.23}}
\multiput(50.43,37.40)(0.12,0.15){15}{\line(0,1){0.15}}
\multiput(52.16,39.64)(0.13,0.12){16}{\line(1,0){0.13}}
\multiput(54.31,41.49)(0.21,0.11){12}{\line(1,0){0.21}}
\multiput(56.79,42.87)(0.34,0.11){8}{\line(1,0){0.34}}
\multiput(59.49,43.72)(0.95,0.10){3}{\line(1,0){0.95}}
\put(68.67,30.00){\line(1,0){14.67}}
\put(55.33,30.00){\line(-1,0){14.67}}
\put(65.00,38.00){\line(1,1){11.00}}
\put(58.33,38.00){\line(-1,1){10.67}}
\put(67.67,24.00){\line(1,-1){11.00}}
\put(60.00,11.00){\circle*{0.67}}
\put(53.33,12.67){\circle*{0.67}}
\put(68.00,12.33){\circle*{0.67}}
\put(48.00,18.00){\circle*{0.67}}
\put(59.67,36.67){\line(1,1){4.00}}
\put(64.67,41.67){\line(1,1){4.00}}
\put(69.67,46.67){\line(1,1){4.00}}
\put(77.33,50.33){NS 5-brane 1}
\put(86.67,30.00){NS 5-brane 2}
\put(81.00,13.00){NS 5-brane 3}
\put(44.67,50.67){NS 5-brane $N_f$}
\put(51.33,5.33){Mirror theory}
\end{picture}
\end{center}

The above information may be completely encoded in the "quiver" diagram 
\cite{DM}. For A and B models the "quiver" diagrams are as follows
\vspace*{2cm}
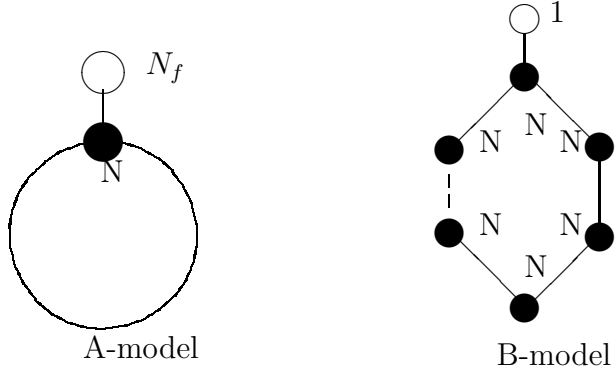
\begin{center}
\unitlength 1mm
\linethickness{0.4pt}
\begin{picture}(102.33,60.67)
\multiput(34.33,42.49)(0.86,-0.09){3}{\line(1,0){0.86}}
\multiput(36.91,42.23)(0.35,-0.11){7}{\line(1,0){0.35}}
\multiput(39.38,41.43)(0.20,-0.12){11}{\line(1,0){0.20}}
\multiput(41.63,40.15)(0.13,-0.11){15}{\line(1,0){0.13}}
\multiput(43.56,38.42)(0.12,-0.16){13}{\line(0,-1){0.16}}
\multiput(45.10,36.34)(0.12,-0.26){9}{\line(0,-1){0.26}}
\multiput(46.18,33.98)(0.11,-0.51){5}{\line(0,-1){0.51}}
\put(46.74,31.45){\line(0,-1){2.59}}
\multiput(46.78,28.86)(-0.10,-0.51){5}{\line(0,-1){0.51}}
\multiput(46.27,26.32)(-0.11,-0.26){9}{\line(0,-1){0.26}}
\multiput(45.26,23.94)(-0.11,-0.16){13}{\line(0,-1){0.16}}
\multiput(43.77,21.82)(-0.13,-0.12){15}{\line(-1,0){0.13}}
\multiput(41.88,20.04)(-0.18,-0.11){12}{\line(-1,0){0.18}}
\multiput(39.67,18.70)(-0.31,-0.11){8}{\line(-1,0){0.31}}
\multiput(37.22,17.84)(-0.86,-0.11){3}{\line(-1,0){0.86}}
\multiput(34.65,17.51)(-1.29,0.10){2}{\line(-1,0){1.29}}
\multiput(32.07,17.71)(-0.36,0.10){7}{\line(-1,0){0.36}}
\multiput(29.59,18.44)(-0.21,0.11){11}{\line(-1,0){0.21}}
\multiput(27.30,19.67)(-0.14,0.12){14}{\line(-1,0){0.14}}
\multiput(25.32,21.34)(-0.11,0.15){14}{\line(0,1){0.15}}
\multiput(23.73,23.39)(-0.11,0.23){10}{\line(0,1){0.23}}
\multiput(22.60,25.71)(-0.11,0.42){6}{\line(0,1){0.42}}
\put(21.97,28.23){\line(0,1){2.59}}
\multiput(21.87,30.82)(0.11,0.64){4}{\line(0,1){0.64}}
\multiput(22.30,33.37)(0.12,0.30){8}{\line(0,1){0.30}}
\multiput(23.26,35.78)(0.12,0.18){12}{\line(0,1){0.18}}
\multiput(24.69,37.94)(0.12,0.11){16}{\line(1,0){0.12}}
\multiput(26.53,39.76)(0.18,0.12){12}{\line(1,0){0.18}}
\multiput(28.71,41.16)(0.30,0.11){8}{\line(1,0){0.30}}
\multiput(31.13,42.08)(0.80,0.10){4}{\line(1,0){0.80}}
\put(34.33,44.33){\line(0,1){5.00}}
\put(34.33,51.67){\circle{6.15}}
\put(34.33,42.33){\circle*{5.20}}
\put(40.00,51.33){$N_f$}
\put(34.00,37.00){N}
\put(31.67,13.67){A-model}
\put(90.33,20.00){\line(1,1){10.00}}
\put(100.33,41.33){\line(0,-1){11.33}}
\put(100.33,41.67){\line(-1,1){10.00}}
\put(90.33,51.33){\line(-1,-1){10.00}}
\put(90.67,20.33){\line(-1,1){10.00}}
\put(100.33,41.67){\circle*{4.00}}
\put(100.33,29.67){\circle*{4.00}}
\put(90.33,20.33){\circle*{4.00}}
\put(90.33,51.00){\circle*{4.00}}
\put(80.33,41.33){\circle*{4.00}}
\put(80.33,30.33){\circle*{4.00}}
\put(90.33,52.67){\line(0,1){4.00}}
\put(90.33,58.67){\circle{4.00}}
\put(80.33,38.00){\line(0,-1){1.67}}
\put(80.33,35.33){\line(0,0){0.00}}
\put(80.33,35.33){\line(0,0){0.00}}
\put(80.33,35.33){\line(0,0){0.00}}
\put(80.33,35.33){\line(0,0){0.00}}
\put(80.33,35.33){\line(0,-1){1.67}}
\put(90.33,43.33){N}
\put(95.00,41.00){N}
\put(84.33,41.00){N}
\put(84.33,30.00){N}
\put(95.00,30.00){N}
\put(90.33,24.33){N}
\put(93.67,58.33){1}
\put(86.67,12.67){B-model}
\end{picture}
\end{center}

In the mirror theory $x^6$ corresponds to the gauge coupling and $\eta_i$'s
are the Fayet-Iliopoules terms. In order to have Higgs branch in the A-model,
the mass of adjoint matter should be zero, in the B-model it means that
$\sum \eta_i=0$. Note that we can have a theory
as the same of B-model but without the fundamental hypermultiplet, it can do
if we do not use NS 5-brane in the A-model or assume that we are far from
it, in this case the fundamental matter decoupled from the theory.  

By the same method as above we can find more complicated models which become
equivalent in the strong coupling limit. These models can be indicated by 
thir "quiver"
diagrams. Note that two theories which become equivalent in strong coupling,
correspond to mirror pairs in three dimensions \cite{BA}. These
mirror pairs of the three dimensional theories have been studied in 
\cite{HO}.
So, exactly the same quiver diagrams occur here too. Note also that 
there are some
difficulties in the case of Sp gauge groups as
we do not understand the action of $S$-duality on the O-planes completely.
\section{Relation with Matrix theory}

In this section we will consider some particular two dimensional 
supersymmetric gauge theories and construct their brane configurations.
These theories have recently been studied in the context of Matrix theory 
formulation of NS 5-brane in Type II string theories.

First consider $N=(4,4)$ $D=2$ supersymmetric gauge theory with gauge 
group $F=U(1)^k$ and $k$ hypermultiplets in the bi-fundamental representation 
of the gauge group ((1,1) representation of $ U(1)_i\times U(1)_{i+1}$
with identification $i\sim i+k$). It has the following brane configuration:

$k$ NS 5-branes at the same point in the directions $x^7,x^8,x^9$ 
and at $k$ points in the $x^6$ direction. We assume that the $x^6$ direction 
is compact with radius $R_6$ and $k$ NS 5-branes positioned in the $x^6$ 
direction in an $Z_k$ invariant way. There is also a D 2-brane wrapped 
around the $x^6$ direction. This D 2-brane is broken into segments 
stretched between each pair of adjacent NS 5-branes.

The freedom on adding an 
arbitrary constant phase to the position of the branes on the circle, leads 
to the decoupling of the $U(1)$ from the gauge group, so that the non-trivial
gauge group is actually $F/U(1)$. The ends of 
the D 2-branes can move inside the NS 5-branes. The position of the D 2-brane
on the NS 5-brane can be interpreted as the scalars in the vector multiplets.
So they parametrize the Coulomb branch of the theory. It has $4k$ dimensions.
Note that since the gauge group is Abelian there is no symmetry breaking in
the Coulomb branch.

If D 2-branes meet from two side of one NS 5-brane, they can reconnect, so we
can have closed D 2-branes which is wrapped around the $x^6$ direction. 
In this case the D 2-brane can leave the NS 5-branes and move far from them
along the $x^7,x^8,x^9$ directions. This is a transition from the Coulomb 
branch
to the Higgs branch. Note that, the Higgs branch occurs via expectation
value of the bi-fundamental hypermultiplets. 

This model is exactly the theory of D 1-brane probe of Type II B string 
theory on the ALE space. To see this, consider the limit of 
$R\rightarrow 0$; in this case, T-duality on the $x^6$ directions
maps us to the Type II B theory on a circle with radius $R_B={1\o {M_s^2 R}}$. 
The $k$ NS 5-branes of Type II A theory, which are located on the transverse 
circle $R$, are mapped into $k$ Kaluza-Klein monopoles in the Type II B string
theory. The $k$ Kaluza-Klein monopoles can be constructed from four 
dimensional multi-Taub-NUT metric tensored with flat space. The non-trivial
metric on the $R^3\times S^1$ is
\be
ds^2=V(x) d{\vec x}^2+V(x)^{-1} (d\theta+{\vec A}\cdot d{\vec x})^2,
\ee
where 
\be
\bigtriangledown V=\bigtriangledown \times {\vec A},\,\,\,\,\,
V=1+R_B \sum_{i=1}^{k}{1\o{|{\vec x}-{\vec x}^i|}}
\ee

The position of the $k$ branes are given by the ${\vec x}^i$ and the angular
variable $\theta$ which has the period proportional to $R_B$. When all the 
branes are separated the space is smooth. For the $k\geq 2$ coalescing branes
the multi-Taub-NUT has an $A_{k-1}$ singularity at the position of the branes.
In the limit $R_B \rightarrow \infty$ the space becomes $R^4/Z_k$. So in our 
case we end up with the Type II B  on the $A_{k-1}$ ALE space. 
Also the D 2-brane maps to D 1-brane and then one can consider, after 
T-duality, motion of the D 1-brane in the ALE space. So 
the theory corresponds to D 1-brane probe of Type II B string theory on the
$A_{k-1}$ singularity which is discussed in \cite{MOR},
where the same theory was discussed. The Higgs branch of the theory 
corresponds to motion of D 1-branes in the ALE space as a background of
the Type II B string theory.  One can also consider the theory with
gauge groups $U(N)^k$. In this case we can study the NS 5-branes in the Type 
II A or Type II B on an $A_{k-1}$ singularity.

Now Consider the theory with gauge group $U(N)^k$ and $k$ bi-fundamental
hypermultiplets. This is exactly the B-model discussed in the previous section,
in the limit that the fundamental hypermultiplet decouples. Moreover we assume 
that the theory is compactified over a circle with radius $R_1$. This theory
corresponds to the matrix theory description of the Type II A string theory
on an $A_{k-1}$ singularity\cite{DM} \cite{MDO}.

In our brane realization, it corresponds to $k$ wrapped NS 5-branes 
(wrap around $x^1$) at the same points in the directions $x^7,x^8,x^9$ and at
$k$ points on the $x^6$ direction in an $Z_k$ invariant way. There are 
also N D 2-branes wrapped around $x^1$ and $x^6$. In fact we can also have 
one D 4-brane, but we are far from it; so it decoupled from the theory. So 
we have exactly the same gauge group and matter content as above. 
These N D 2-branes can break into segments stretched between each pair of 
adjacent NS 5-branes. 
 
This theory has a Coulomb branch as well as a Higgs branch. The Higgs branch
can be obtained from the expectation value of the bi-fundamental 
hypermultiplets 
and describes the bulk theory. The Coulomb branch describes the decoupled 
six dimensional theory.
It has $4kN$ dimensions with tube-like metric. In  fact this theory
corresponds to the Type II A Matrix theory on an ADE singularity as studied
in\cite{SS}\cite{MDO}.

Let us apply the $U$-transformation to this configuration as before.
Doing so, we will have N D 2-branes wrapped around $x^1, x^2$, and $k$
wrapped D 4-branes at points on $x^6$.\footnote{Note that the decoupled  
fundamental matter maps to NS 5-brane wrapped around $x^1$ direction and
in general can intersect the D 2-branes and from point of 
view of two dimensional theory, it gives us the matter in the adjoint 
representation}
Since both directions of wrapped 
D 2-branes are compactified, and in general they can be in the same order, so
in fact we have 2+1 dimensional  gauge theory with gauge group $U(N)$ 
compactified on the torus. This gauge theory has 16 supercharges and is the 
bulk
theory; but presence of the $k$ D 4-branes which can be interpreted as 
impurities in the 2+1 dimensional gauge theory, break half of the
supercharges. These impurities correspond to the $k$ hypermultiplets in the 
fundamental representation. 

The bulk theory which is described by the D 2-branes, has seven scalars 
in the
adjoint representation corresponding to the seven directions transverse to D 
2-branes
and has $SO(7)$ global symmetry. But presence of the D 4-branes break this
symmetry to $SO(4)\times SU(2)$ and the scalars break in to two parts. 
One, the
scalars in the vector multiplet which parametrize the Coulomb branch 
(fluctuation of D 2-branes in the directions $x^2,x^3,x^4,x^5$) and another,
the scalars in the hypermultiplets which parametrize the Higgs branch 
(fluctuation of D 2-branes in the directions $x^7,x^8,x^9$). The $SO(4)$  
acts on the scalars in the vector multiplet and $SU(2)$ on the scalars in the
hypermultiplets.

Here, the bulk theory, which is the space-time physics with background 
$k$ Type
II B wrapped NS 5-branes, can be described by the Coulomb branch and the 
Higgs branch describes the decoupled theory. In the limit of $R_6 \rightarrow
0$, this theory is 1+1 dimensional $N=(4,4)$ theory with gauge group $U(N)$ 
and one adjoint hypermultiplet and $k$ fundamental hypermultiplets.

The decoupled theory is described by the Higgs branch of the theory. In this 
case the $k$ hypermultiplets get expectation values and N D 2-branes break 
into 
segments stretched between D 4-branes. The Higgs branch is parametrized by
the fluctuation of the D 2-branes in the directions $x^7,x^8,x^9$ and 
the last scalar corresponds to non-zero component of the gauge field
$A_6$. This branch has $4kN$ dimensions, as we expect from the mirror theory.
 If we go to the strong coupling limit, the last scalar will be the 
fluctuation of D 2-branes in the $11^{th}$ direction and as we said, in this
limit, it leads to R-symmetry enhancement from $SU(2)$ to $SO(4)$
\cite{SS}\cite{BA}. As we said there is a Higgs branch for the case
$k=1$, it means that the theory for one NS 5-brane should also
be non-trivial at the decoupled limit !

Note that it seems that, the two dimensional mirror symmetry which we 
considered
here, corresponds to the duality between Type II A on an $A_{k-1}$ singularity
and $k$ NS 5-branes in Type II B and also the duality between Type II B on 
an $A_{k-1}$ singularity
and $k$ NS 5-branes in Type II A. Note also that, This mirror symmetry may 
shed light on the dual realization of the 1+1 dimensional $N=(4,4)$ gauge 
theory in the such way that the tube metric is absent, but the dual theory
has ADE singularity as conjectured in \cite{SOWI}\cite{DS} and argued in
\cite{VAFA} and recently discussed in the second reference in  
\cite{SS}.

\section{Acknowledgements}

I would like to thank F. Ardalan, M. H. Sarmadi for useful discussions.

\end{document}